\documentclass[fleqn,usenatbib]{mnras}
\usepackage{amsmath}
\usepackage{graphicx}
\usepackage{amssymb,txfonts}
\usepackage{xspace}

\newcommand{\msun}{{\rm M}_{\sun}}

\newcommand{\xte}{{\textit{RXTE}}\xspace}

\newcommand{\xmm}{{\textit{XMM-Newton}}\xspace}

\newcommand{\nustar}{{\textit{NuSTAR}}\xspace}
\newcommand{\suzaku}{{\textit{Suzaku}}\xspace}

\newcommand{\source}{{GX~339--4}\xspace}

\title[Disc truncation in the hard state of GX 339--4]
{Comparison of spectral models for disc truncation in the hard state of GX 339--4}
\author[M. A. Dzie{\l}ak et al.]
{Marta A. Dzie{\l}ak,$^1$\thanks{mdzielak@camk.edu.pl, aaz@camk.edu.pl, mitsza@camk.edu.pl}, 
Andrzej A. Zdziarski,$^{1\star}$ Micha{\l} Szanecki,$^{1\star}$ Barbara De Marco,$^{1}$\newauthor Andrzej Nied{\' z}wiecki$^{2}$ and Alex Markowitz$^{1,3}$\\
$^1$Nicolaus Copernicus Astronomical Center, Polish Academy of Sciences, Bartycka 18, PL-00-716 Warszawa, Poland\\
$^2$Department of Astrophysics, {\L}{\'o}d{\'z} University, Pomorska 149/153, 90-236 {\L}{\'o}d{\'z}, Poland\\
$^3$University of California, San Diego, Center for Astrophysics and Space Sciences, 9500 Gilman Dr, La Jolla, CA 92093-0424, USA}
\begin{document}


\date{Accepted 2019 March 4. Received 2019 March 4; in original form 2018 November 21}

\pagerange{\pageref{firstpage}--\pageref{lastpage}} \pubyear{2018}

\maketitle

\label{firstpage}

\begin{abstract}
We probe models of disc truncation in the hard spectral state of an outburst of the well-known X-ray transient GX 339--4. We test a large number of different models of disc reflection and its relativistic broadening, using two independent sets of codes, and apply it to a \textit{Rossi X-ray Timing Explorer\/} spectrum in the rising part of the hard state of the 2010/11 outburst. We find our results to be significantly model-dependent. While all of the models tested show best-fits consistent with truncation, some models allow the disc to extend close to the innermost stable circular orbit (ISCO) and some require substantial disc truncation. The different models yield a wide range in best-fit values for the disc inclination. Our statistically best model has a physical thermal Comptonization primary continuum, requires the disc to be truncated at a radius larger than or equal to about two ISCO radii for the maximum dimensionless spin of 0.998, and predicts a disc inclination in agreement with that of the binary. Our preferred models have moderate Fe abundance, $\gtrsim$2 times solar. We have also tested the effect of increasing the density of the reflecting medium. We find it leads to an increase in the truncation radius, but also to an increase in the Fe abundance, opposite to a previous finding. 
\end{abstract}

\section{Introduction}
\label{intro}

\vbox to 0.5cm{\vfill}

The standard model \citep{ss73,nt73} of accretion onto black holes (BHs) predicts formation of a viscously dissipating optically-thick disc extending down to the radius of the innermost stable circular orbit, $R_{\rm ISCO}$. This model explains well soft states of accreting systems, e.g., the soft spectral state of BH X-ray binaries. However, it cannot explain states in which BH binaries and active galactic nuclei emit predominantly hard X-ray radiation, see e.g., a review by \citet*{done07}. That radiation has to be instead emitted by some hot plasma. The location of this plasma remains poorly understood. 

Among accreting BH sources, low-mass BH X-ray binaries represent an important class. They are transient in most of the known cases \citep*{coriat12}, and spend most of the time in a quiescent state. In that state, the optically-thick disc is predicted by the disc instability model to have a large inner truncation radius, $R_{\rm in}\sim 10^4 R_{\rm g}$ \citep*{lasota96,dubus01}, where $R_{\rm g}\equiv GM/c^2$, and $M\sim 10\msun$ is the BH mass. The quiescence truncation radius has been measured in the case of the low-mass BH X-ray binary V404 Cyg to be $R_{\rm in}\ga 3.4\times 10^4 R_{\rm g}$ (\citealt{bernardini16}; see \citealt*{narayan97} for an earlier estimate). During quiescence, matter transferred from the companion accumulates in the disc and its inner radius continuously decreases; see, e.g., fig.\ 13 of \citet{dubus01}. Confirming the prediction of the decrease in $R_{\rm in}$, an upper limit of $R_{\rm in}\la 1.2\times 10^4 R_{\rm g}$ in V404 Cyg was obtained 13 h before the onset of the 2015 X-ray outburst \citep{bernardini16}. 

For the parameters of \citet{dubus01}, the hydrogen-ionisation instability triggering the outburst starts at a radius of $\sim\! 10^{10}$ cm ($\sim\! 10^3 R_{\rm g}$), a radius slightly larger than $R_{\rm in}$ at that point of time, $\approx 6\times 10^9$ cm. These radii are much lower than the typical outer disc radius of $\sim\! 10^{11}$ cm, and this type of outburst is called 'inside-out'. The truncation radius at the onset of an outburst can be estimated using observed time delays between the onsets of optical and X-ray flux rises as  compared to the difference of the viscous time scales, $t_{\rm vis}=R^2/\nu$ (where $\nu$ is the kinematic viscosity), at the inner disc radius at the optical flux rise, $R_{\rm in}(V)$, and the X-ray one, $R_{\rm in}(X)$, with the latter assumed by \citet{dubus01} to be $5\times 10^8$ cm ($\sim 300 R_{\rm g}$). In V404 Cyg, a $\sim$7-d lag has been observed by \citet{bernardini16}, from which they derived $R_{\rm in}(V)\approx 0.9$--$2.2\times 10^9$ cm ($\sim 10^3 R_{\rm g}$), somewhat lower than the value of \citet{dubus01} (and much lower than the observational upper limit obtained 13 h before the outburst, see above). An almost identical optical-to-X-ray delay, $\approx 7\pm 1$ d, was found in a new BH transient, ASASSN-18ey, by \citet{tucker18}, who found $R_{\rm in}(V)\approx 0.8$--$2.5\times 10^9$ cm.

Since the viscous time scale decreases with decreasing radius (as $\propto R^{1/2} T_{\rm c}^{-1}$, where $T_{\rm c}$ is the midplane disc temperature, increasing with decreasing $R$), the assumption of the onset of X-ray outburst at $R_{\rm in}\approx 5\times 10^8$ cm has only a minor effect on the derived value of $R_{\rm in}(V)$. On the other hand, if the viscous build-up of the disc continued after the onset of the X-rays, the disc should have reached $R_{\rm ISCO}$ ($\sim\!\! 10^6$--$10^7$ cm) within a couple more days, i.e., at the beginning of the hard spectral state. We note, however, that our understanding of the mechanisms of truncation, both in quiescence and in outburst, remains very limited. The quiescence truncation radius in \citet{dubus01} is determined by their equation (14), from \citet{menou00}, which, as they note, is both phenomenological and uncertain. The uncertainty is caused by our poor understanding of disc evaporation mechanisms. In the hard state, a number of different mechanisms to keep the disc truncated and change $R_{\rm in}$ with changing luminosity and source history (including a hysteretic behaviour of state transitions) have been proposed, e.g., \citet*{meyer05}, \citet{petrucci08}, \citet{begelman14}, \citet{kylafis15}, \citet{cao16}. We conclude that the correct theoretical dependence of $R_{\rm in}$ on the time and luminosity after the onset of an outburst in the hard and intermediate states remains unknown.

Indeed, there is abundant evidence for the disc not reaching the ISCO at the beginning of the hard state. For example, \citet{mcclintock01} and \citet{esin01} found strong evidence for $R_{\rm in}\ga 50 R_{\rm g}$ three weeks after the beginning of the 2000 outburst of the BH binary XTE J1118+480, at a luminosity of $\sim\! 10^{-3}$ of the Eddington luminosity, $L_{\rm E}$. On the other hand, $R_{\rm in}$ certainly equals $R_{\rm ISCO}$ (or it is very close to it) during the soft spectral states \citep*{ebisawa91, ebisawa93, gierlinski04, steiner11}. Thus, we know that $R_{\rm in}$ has to decrease from some hundreds or $\sim 10^3 R_{\rm g}$ at the onset of the disc instability to $R_{\rm ISCO}$ at the onset of the soft state. 

In the case of \source, a well-studied transient low-mass BH X-ray binary, a number of authors found the presence of an optically-thick disc already extending very close to the ISCO in the hard spectral state at luminosities $\ga 0.01 L_{\rm E}$ (\citealt*{miller06, miller08, reis08,reis10,tomsick08, petrucci14,furst15, garcia15}, hereafter G15; \citealt{wang_ji18b}). The method used was X-ray reflection spectroscopy, in which theoretical reflection spectra are fitted to data. On the other hand, other authors found highly truncated discs in the same state using the same method, and, in some cases, using the same observations \citep*{done10, kolehmainen14,plant15,basak16}. A truncated disc can also explain the relatively long reverberation lags measured in the soft X-ray response of the inner disc to variability of hard X-rays \citep*{demarco15, demarco17,mahmoud19}. 

\begin{figure*}
\centering
\includegraphics[width=12.93cm]{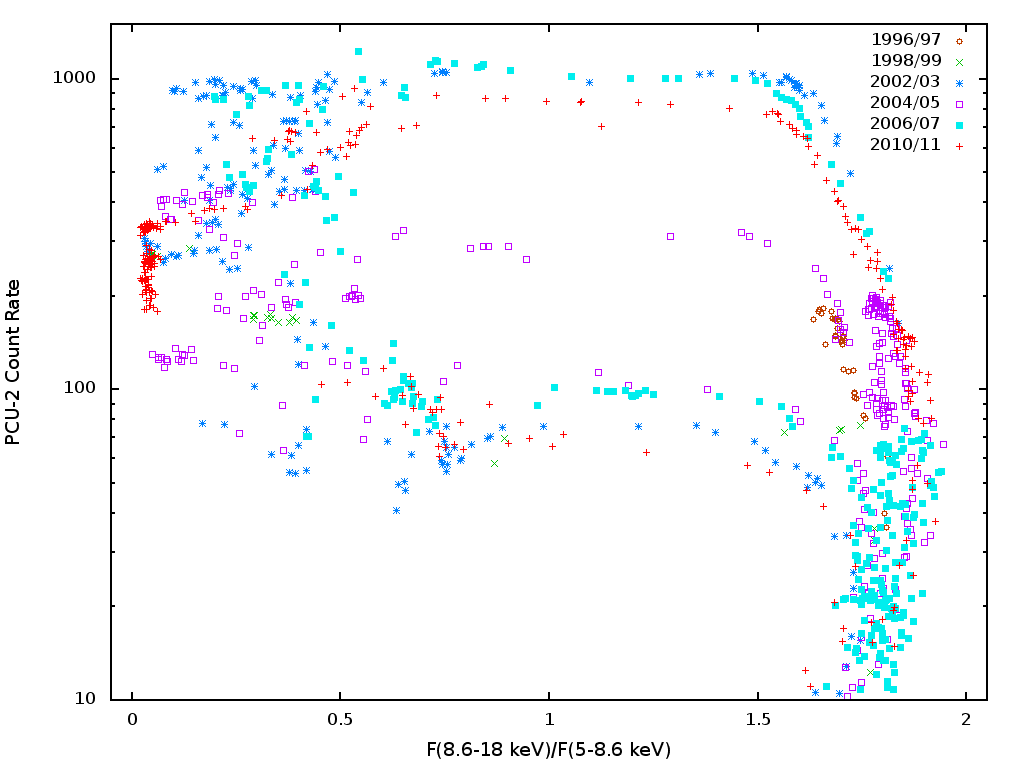}
\caption{The diagram of the 3--45 keV count rate vs.\ hardness for 1107 PCA observations of \source during its six major outbursts, identified by different symbols and colours. The hardness is defined as the energy flux ratio of the 8.6--18 to 5--8.6 keV bands. }
\label{r_h}
\end{figure*}

Some of the findings of the disc extending close to the ISCO in the luminous hard state of \source were attributed to the effect of instrumental pile-up, present in some of the \xmm fitted spectra (as discussed by \citealt{done10, kolehmainen14,basak16}). This effect, present in CCD detectors, is absent in the data from \nustar, analysed by \citet{furst15} and \citet{wang_ji18b}. However, while \citet{furst15} advocated a low truncation radius, a fraction of their best-fitting models had large truncation radii, and thus their results cannot be considered as conclusive. Then, while \citet{wang_ji18b} found $R_{\rm in}\sim R_{\rm ISCO}$ in their models, their observations were taken in a low-luminosity state of \source, which limited the statistical significance of those results. The pile-up effect was also absent in the analysis of G15, who used data from the Proportional Counter Array (PCA) on board the {\it Rossi X-ray Timing Explorer}\/ (\xte). G15 used the correction to the spectral response of the PCA of \citet{garcia14b} and combined hard-state PCA observations with similar X-ray fluxes, which resulted in well-calibrated spectra with very large number of counts. Then, in spite of the limited spectral resolution of the PCA (typical for proportional counters), detailed reflection spectra could be fitted. The reflection model used by G15 was then extended by \citet{steiner17} (hereafter S17), who included the effect of scattering of the reflection photons in the hot plasma emitting the X-rays irradiating the disc. The works of G15 and S17 appear to represent some of the best documented cases for low truncation radii in the hard state of \source. However, motivated by the wealth of existing controversial results, we have embarked on an independent analysis of the PCA spectra in the hard state of \source. 

\source was discovered in 1972 \citep{markert73}. It has since then been the most often outbursting transient BH binary. The distance to the source, $D$, is relatively uncertain. \citet{z04} obtained $7\,{\rm kpc}\la D\la 9\,{\rm kpc}$. \citet{heida17} found $D\ga 5$ kpc and their preferred value was $D\approx 9$ kpc.  Its mass function is $1.91\pm 0.08\msun$ and the mass ratio is $0.18\pm 0.05$, which, with a constraint on the donor mass, give $M \la 9.5\msun$ \citep{heida17}. On the other hand, \citet{parker16} found $D\approx 8$--10 kpc and $M\approx (8$--$12)\msun$ (including both the statistical and systematic errors) based on their X-ray spectral fits. In our estimates of the luminosity and the Eddington ratio, we assume $D=8$ kpc and $M= 8\msun$ (as well as we assume the hydrogen fraction of 0.7 in the value of the Eddington luminosity). The binary inclination, $i_{\rm b}$, of \source has been constrained by \citet{heida17} to be $37\degr\leq i_{\rm b}\leq 78\degr$. We note that an inner part of the disc can be aligned with the BH rotation axis, which, in turn, can be misaligned with the binary axis. Thus, the inclination derived from X-ray fits, $i$, does not have to necessarily equal $i_{\rm b}$ if the disc extends close to the ISCO.

\section{The data reduction}
\label{data}

\xte was an X-ray observatory launched in 1996 and operational until 2012. The PCA \citep{jahoda06} was one of the two pointed detectors on board \xte. The PCA operated in the nominal energy range of 2--60 keV, and it conducted about 1400 pointed observations of \source.

We have analysed the PCA observations of \source and extracted PCA spectra\footnote{As described in the \xte data reduction cookbook: \url{http://heasarc.nasa.gov/docs/xte/recipes/cook_book.html}.}. We used only the Proportional Counter Unit (PCU) 2, since it was consistently operational during {\xte}'s entire lifetime, as well as the best-calibrated unit. We used all of the layers of the detector. We have applied the standard dead-time correction, using the commands {\sc pcadeadcalc2} and {\sc pcadeadspect2}. For each observation, we used the PCA background model for bright sources. Using the X-Ray Spectral Fitting Package ({\sc xspec}) v.~12.10.0c \citep{arnaud96}, we have fitted absorbed power-law models to all the spectra, and defined the spectral hardness as the ratio of the energy fluxes (from the model) between the 8.6--18 and 5--8.6 keV photon energy ranges. We show the PCA count rate as a function of that hardness in Fig.\ \ref{r_h}, where we also identify the six major outbursts of \source observed by \xte. 

\begin{figure}
\centering
\includegraphics[width=8cm]{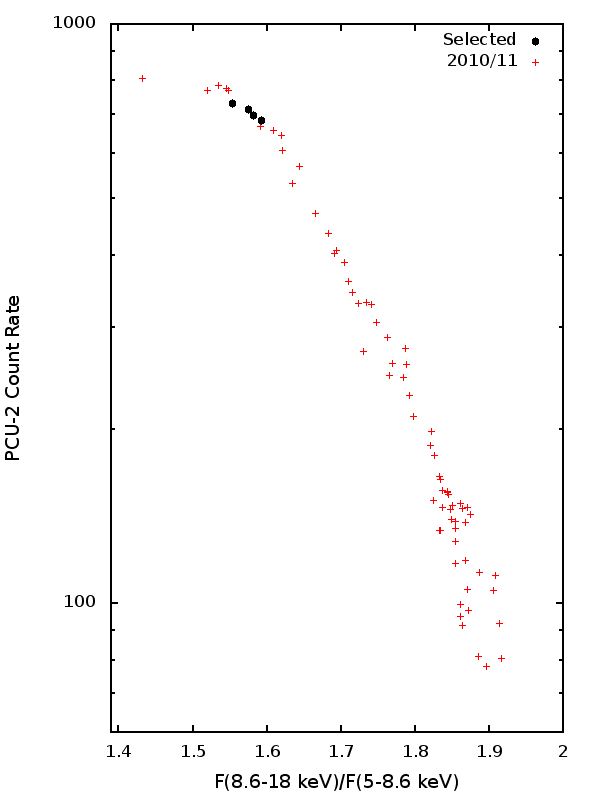}
\caption{The 3--45 keV count rate vs.\ hardness for the PCA observations of \source during the rise of the 2010/11 outburst (which is a subset of the data shown in Fig.\ \ref{r_h}). The black circles mark the observations forming the average spectrum fitted in this work. }
\label{groups}
\end{figure}

Then, in order to increase the statistical precision of spectral fits, we have combined spectra with similar flux and hardness. We have chosen the 2010/11 outburst, since it has the coverage of the rising part of the hard state with the highest dynamic range among all of the observed outbursts. The analysed observations are shown in Fig.\ \ref{groups}. We started from the bottom, and included observations with increasing fluxes until a minimum of $10^7$ counts was accumulated. In this way, we have obtained 10 sets of spectra in the hard state.  We then added together observations within a set in two ways. In one, we followed the method proposed by G15. Every observation in an averaged set was fitted by an absorbed power-law. The best fit parameters within each set were averaged and used as input parameters in {\sc fakeit} command of {\sc xspec}, in order to simulate an average spectrum for the continuum. The residuals were summed together, and the result added to the average absorbed power-law spectrum. For the response file, we used one of the original responses for each combined set, and we have tested that a given choice has no effect on the results. We ended up with 10 spectra at different luminosities in the hard state, which we denote using the letter G. In the second method, we obtained an summed spectrum for each set using the standard routine {\sc addspec} included in the {\sc ftools} package, which routine also generates the appropriate response file. The resulting spectra are denoted with the letter A. In both cases, we applied the correction to the PCA effective area of \citet{garcia14b}, {\tt pcacorr}, and, following the recommendation of that work, added a 0.1 per cent systematic error. As shown in Section \ref{g15}, the two procedures yield similar spectra and fitting results. 

We have performed extensive spectral fitting with more than 10 different models for most of the spectral groups and for each of the averaging methods. We have generally found a strong model dependence of the results, and our full set of results has therefore become very complex. We have therefore decided to focus the present work on showing and discussing results obtained from the analysis of only one spectrum, and defer the analysis of all the others as well as of the evolution of spectral parameters during the outburst to a follow-up paper. The chosen spectrum includes four among the brightest observations of the rising part of the outburst, see the black points in Fig.\ \ref{groups}. The next brighter average spectrum is already located on an approximately horizontal part of the hardness-count rate diagram (together with $\sim$10 later observations of that outburst, see Fig.\ \ref{r_h}) and thus is likely to have different spectral properties than those characterizing the rising part. We have also checked that all of the four individual observations in the chosen average spectrum, taken on 2010 April 2, 3, 4 and 5 (Obsid 95409-01-13-03, 95409-01-13-00, 95409-01-13-04, 95409-01-13-02), have similar residuals with respect to the power-law fits. This is, e.g., not the case for the previous, fainter, spectrum, where one long observation shows much stronger residuals with respect to the fitted power-law than the remaining observations (possibly a consequence of calibration issues). Our selected spectrum has $1.01\times 10^7$ counts. Its average count rate is 705.6 s$^{-1}$. It is similar in flux and hardness to spectrum B of G15 (note that the count rates given in G15 have been adjusted for the detector gain change over the lifetime of \xte and thus cannot be directly compared with our count rate). Our spectrum has also relatively similar flux and hardness to \xmm/EPIC-pn spectrum 7, taken during the same outburst on 2010 March 28 \citet{basak16}.

\section{Spectral analysis}
\label{spectral}

\subsection{Methodology}
\label{method}

After preliminary spectral fitting with {\sc xspec}, we used the command {\sc steppar} to scan the parameter space and find the model with the overall lowest $\chi^2$. After that, we determine the 90 per cent confidence range for a single parameter, $\Delta\chi^2=+ 2.71$, corresponding to the value farthest away from the best fit, also using {\sc steppar}. Note that occasionally this procedure gives a limit within a local minimum separated from the best fit by a parameter range with $\Delta\chi^2> 2.71$.

We then check for degeneracy between parameters of our models and further explore their parameter spaces using a Markov Chain Monte Carlo (MCMC) algorithm. We use the {\tt xspec\_emcee} implementation (by Jeremy Sanders based on \citealt{foreman13} and \citealt{goodman10}). Also, we have used the standard {\sc xspec} tools, for which we have tested both types of algorithms implemented, Metropolis-Hastings and Goodman-Weare, and two assumed error distributions, Gaussian and Cauchy. We have applied the MCMC method to all models in this paper, but present results graphically only for model 6, which is as presented in Section \ref{Compton} below. For that, we used {\tt xspec\_emcee} with 50 so-called walkers with 240,000 iterations each, discarding the first 5,000. The auto-correlation length calculated for each free fit parameter is approximately 80 times smaller than chain length. We note that with the MCMC we have not found any better fits than those found using {\sc steppar}. We also note that the confidence ranges determined with the MCMC method in {\sc xspec} usually correspond to the global minima only, and do not include possible local minima away from the global one (which we find using {\sc steppar}).

Most of our models are not nested. Thus, we need a criterion to compare models different from the F-test. We use the Akaike information criterion, AIC \citep{akaike73,sugiura78}, which has been relatively widely used in astrophysics (e.g., \citealt{koen06,liddle07,natalucci14,lubinski16,tang18}). We use a formula with a correction for the finite size of sample,
\begin{equation}
\textup{AIC}_i = 2 m -2 C_{\rm L} + \chi^2 + \frac{2 m(m+1)}{n-m-1}, 
\end{equation}
where $C_{\rm L}$ is a likelihood function of the true model (which depends only on the data set), $m$ is the number of free parameters of a model, $n$ is the number of channels in the fitted spectrum, and $i$ is the model number. The lower the AIC value, the better the model. Since the models are compared through a difference in their AIC values, the likelihood $C_{\rm L}$ cancels out. The relative likelihood for a model with a larger value of AIC$_i$ compared to the best one with the minimum AIC, which we denote as AIC$_0$, is $\exp[({\rm AIC}_0-{\rm AIC}_i)/2]$, which is unity for ${\rm AIC}_i={\rm AIC}_0$ and $\approx 0$ for ${\rm AIC}_i\gg {\rm AIC}_0$. Instead of normalizing to the best model, we can also normalize the likelihood to unity for the sum of all considered $I$ models \citep{akaike78},
\begin{equation}
p_k  = \frac{\exp(-{\rm AIC}_k/2)}
{\sum_{i=1}^{I} \exp(-{\rm AIC}_i/2)},
\end{equation}
with $\sum p_k=1$. We give the values of $p_k$ for our models in Table \ref{fits}.

\subsection{The ISM absorption column toward \source}
\label{abs}

In all our models, the ISM absorption is taken into account by the model {\tt tbabs} \citep*{wilms00}. The description of that model\footnote{\url{http://pulsar.sternwarte.uni-erlangen.de/wilms/research/tbabs}} recommends the use of the cosmic abundances of \citet{wilms00}. However, the actual ISM abundances in the direction to \source remain uncertain, and G15 used instead the abundances of \citet{anders89}. In our models, we have tested the effects of changing the abundances on the fits, and found it to be relatively minor, except for the fitted value of the absorption column density, which was substantially higher for the abundances of \citet{wilms00}. Therefore, from this point onwards, we follow G15 and use the abundances of \citet{anders89}.

The actual value of the absorption column towards \source, $N_{\rm H}$, also remains somewhat uncertain. Since the data we use are for $E>3$ keV only, we need to constrain the allowed range of $N_{\rm H}$. \citet{z98} listed a number of previous determinations of it, and found it to be in the approximate range of (5--$7)\times 10^{21}$ cm$^{-2}$. The best-fit models of \citet{furst15} yield $\approx 8\times 10^{21}$ cm$^{-2}$. \citet{basak16} obtained $(7.0\pm 0.1)\times 10^{21}$ cm$^{-2}$ for the abundances of \citet{anders89} when fitting a set of seven \xmm/EPIC-pn observations. Based on the results listed above, we hereafter assume $N_{\rm H}$ to be in the range of (4--$8)\times 10^{21}$ cm$^{-2}$. We stress that since X-ray absorption depends mostly on the column densities of metals, the true value of the ISM $N_{\rm H}$ depends strongly on the abundances of heavy elements.

\subsection{Models with relxill and two Fe abundances}
\label{g15}

{\def\arraystretch{1.4}
\begin{table*}
\caption{The spectral fitting results for our models applied to data set G. Model 0 follows the original assumptions of G15, which appear unphysical, and model 6 is our best model, with a physical primary continuum from thermal Comptonization. All models have the ISM absorption term {\tt tbabs}. Model 0: [\texttt{relxill}(free $Z_{\rm Fe}$)+\texttt{xillver}($Z_{\rm Fe}$=1)]{\tt gabs}; Models 1 and 2: \texttt{relxill+xillver}; Model 3: \texttt{relxillD+xillverD};  Models 4 and 5: \texttt{reflkerrExp+hreflectExp}; Model 6: \texttt{reflkerr+hreflect}. Models 0, 1, 3, 4 and 6 have high and low ionization values for the close and distant reflectors, respectively, while models 2 and 5 have the ionization structure reversed. The effect of scattering of the reflection component is taken into account in separate models with {\tt simplcut} (see Section \ref{spectral}), for which we give here the values of the scattering fraction, $f_{\rm sc}$, in the last row of the table. 
}
\begin{tabular}{cccccccc}
\hline
Parameter/Model & 0 & 1 & 2 & 3 & 4 & 5 & 6 \\
\hline
\hline
$N_{\rm H}/10^{21}\,{\rm cm}^{-2}$ & $5.2^{+1.8}_{-1.2}$ & $4.7^{+1.5}_{-0.3}$ & $6.5^{+1.3}_{-1.7}$ & $6.1^{+0.7}_{-0.6}$ & $4.4^{+1.9}_{-0.4}$ & $6.4^{+0.8}_{-1.1}$ & $4.3^{+0.5}_{-0.3}$  \\
$\Gamma$ & $1.70^{+0.07}_{-0.04}$ & $1.66^{+0.03}_{-0.04}$ & $1.72^{+0.02}_{-0.03}$ & $1.70^{+0.01}_{-0.05}$ & $1.66^{+0.06}_{-0.02}$ & $1.72^{+0.03}_{-0.01}$ & -- \\
$y$ & -- &  --&  --&-- &-- & -- &$1.19^{+0.05}_{-0.08}$  \\
$E_{\rm cut}$/keV& $200^{+130}_{-50}$ & $250^{+50}_{-20}$ & $300^{+80}_{-50}$ & $300$f & $240^{+50}_{-50}$ & $280^{+50}_{-20}$ & -- \\
$kT_{\rm e}/1\,$keV &  --&  --&-- &-- & -- & --& $20^{+3}_{-2}$  \\
$R_{\rm in}/R_{\rm ISCO}$ & $11_{-10}^{+10}$ & $19^{+33}_{-6}$ & $53^{+\infty}_{-26}$   & $55^{+\infty}_{-34}$ & $15^{+31}_{-12}$ & $58^{+\infty}_{-28}$& $47^{+\infty}_{-45}$   \\
$Z_{\rm Fe}$  & $8.1^{+1.9}_{-5.5}$ & $3.1^{+2.0}_{-0.3}$ & $2.4^{+0.3}_{-0.2}$ & $4.9^{+4.1}_{-0.9}$ & $3.9^{+0.8}_{-1.4}$ & $2.6^{+0.6}_{-0.4}$ & $3.3^{+1.7}_{-1.0}$  \\
$i\,[\degr]$ & $29_{-29}^{+31}$ & $3^{+33}_{-3}$ & $43^{+17}_{-23}$ & $3^{+43}_{-3}$ & $9^{+32}_{-9}$ & $43^{+21}_{-19}$ & $49^{+34}_{-26}$  \\
$\mathcal{R}$ (inner) & $0.059^{+0.001}_{-0.001}$ & $0.170^{+0.004}_{-0.005}$ & $0.144^{+0.004}_{-0.003}$ & $0.059^{+0.033}_{-0.006}$ & $0.25^{+0.04}_{-0.19}$ & $0.35^{+0.06}_{-0.12}$ & $0.42^{+0.36}_{-0.12}$  \\
$\log_{10}\xi$ (inner) & $3.7^{+0.2}_{-0.5}$ &  $3.9^{+0.1}_{-0.1}$ & $0.0^{+2.3}$ & $3.7^{+0.1}_{-0.1}$ & $3.9^{+0.1}_{-0.1}$ & $1.7^{+0.7}_{-1.7}$ & $3.9^{+0.1}_{-0.3}$   \\
$\log_{10}\xi$ (outer) & $0$f &   $1.7^{+0.5}_{-1.7}$ & $3.8^{+0.2}_{-0.3}$  & $0.7^{+1.0}_{-0.3}$ & $2.0^{+0.3}_{-0.4}$ & $3.7^{+0.1}_{-0.3}$ & $0$f  \\
$n_{\rm e}/1\,{\rm cm}^{-3}$ &$10^{15}$f &$10^{15}$f &$10^{15}$f &$10^{19}$f &$10^{15}$f &$10^{15}$f &$10^{15}$f\\ 
$\delta({\tt gabs})$ &  $0.011^{+0.011}_{-0.009}$ &  --&  --&-- &-- & -- & -- \\ 
$kT_{\rm bb}/1\,$keV &  --&  --&-- &-- & -- & -- & $0.34^{+0.04}_{-0.09}$ \\
\hline 
$\chi_\nu^2$  & 65.6/61 & 68.7/61 & 68.3/61 & 72.4/62 & 69.1/61 & 69.2/61 & 62.1/61  \\
$p_i$ (AIC)  & 0.136 & 0.007 & 0.008 & 0.018 & 0.024 & 0.023 & 0.784  \\
\hline
$f_{\rm sc}$ & $0^{+0.27}$ & $0^{+0.29}$ & $0^{+0.31}$ & $0^{+0.12}$ & $0^{+0.26}$ & $0^{+0.79}$ & $0.30_{-0.30}^{+0.15}$  \\
\hline
\end{tabular}\\
{\it Notes:} We assume the dimensionless spin $a_*=0.998$, for which $R_{\rm ISCO}\approx 1.237 R_{\rm g}$. $\delta({\tt gabs})$ is the energy-integrated depth of the 7.2 keV line. We use the symbol $\infty$ to denote that the upper limit of $R_{\rm in}$ approaching $R_{\rm out}=10^3 R_{\rm g}$, which is the maximum radius for which relativistic broadening is calculated in {\tt relxill} and {\tt reflkerr}. The Compton parameter is defined as $y\equiv 4(kT_{\rm e}/m_{\rm e}c^2)\tau_{\rm T}$, where $\tau_{\rm T}$ is the Thomson optical depth of the slab (approximating the corona). 'f' denotes a fixed parameter. The fitted ranges of $N_{\rm H}/10^{21}\,{\rm cm}^{-2}$, $Z_{\rm Fe}$ and $\log_{10}\xi$ are constrained to $[4,\,8]$, $\leq$10 and $\geq 0$, respectively. The reflection fraction, $\mathcal{R}$, is the ratio of photons emitted toward the disc to those escaping to infinity in {\tt relxill} \citep{dauser16}, and is the fraction of locally emitted photons in the direction of the disc in {\tt reflkerr} \citep{niedzwiecki18}.
\label{fits}
\end{table*}}

G15 assumed a reflection emissivity profile of $\propto R^{-3}$ down to $R_{\rm in}\geq R_{\rm ISCO}$. This can corresponds either to a corona sandwiching a standard disc (except for the neglect of the zero-stress inner boundary condition) or a central hot flow irradiating an outer disc. In either case, the assumed profile implies the reflection profile to be moderately centrally dominated. The reflected spectra are relativistically broadened. G15 used version 0.2g of the model {\tt relxill}\footnote{\url{http://www.sternwarte.uni-erlangen.de/~dauser/research/relxill/}} \citep{garcia14a}. That model combines the {\tt xillver} model \citep{garcia10,dauser10}, which describes rest-frame angle-dependent reflected spectra (under the assumption of the constant electron density of the reflecting medium at $n_{\rm e}=10^{15}$ cm$^{-3}$), with the relativistic blurring model of {\tt relline} \citep{dauser10}. The incident photons have an e-folded power-law photon spectrum, $\propto E^{-\Gamma} \exp(-E/E_{\rm cut})$, where $\Gamma$ is the photon index and $E_{\rm cut}$ is the e-folding energy. In addition to the relativistically-blurred reflection, it contains a static-reflection component {\tt xillver}, which accounts for reflection from remote parts of the accretion disc. Also, they found it necessary to include an additional narrow absorption line at $\approx$7.2 keV.  Thus, their model has the form of {\tt tbabs(relxill+xillver)gabs}. Furthermore, G15 found that in order to achieve a good fit with this model, they had to fix that abundance for the static-reflection component ({\tt xillver}) at the solar value while they allowed a free Fe abundance of the relativistically-blurred component. In addition, they assumed that the reflecting surface for the static component is close to neutral, with an ionization parameter $\xi=1$ erg cm$^{-2}$ s$^{-1}$, where 
\begin{equation}
\xi\equiv \frac{4\upi F_{\rm irr}}{n_{\rm e}},
\label{xi}
\end{equation}
and $F_{\rm irr}$ is the irradiating flux in the 13.6 eV--13.6 keV band. That value of $\xi$ is the minimum one for which the above reflection models are defined.
 
We first fit our data set G with this model. However, we use the current version of the {\tt relxill} software, 1.2.0. This version agrees relatively well with the independently developed code, {\tt reflkerrExp}, of \citet*{niedzwiecki18}. 

We found that we cannot constrain the BH spin, with no difference in $\chi^2$ between the maximum spin of $a_*=0.998$ and 0. Therefore, we hereafter fix $a_*=0.998$, for which $R_{\rm ISCO}\approx 1.237 R_{\rm g}$. Also, we assume the largest outer radius allowed in the {\tt relxill} model, $R_{\rm out}=10^3 R_{\rm g}$. For the spectrum obtained with the method of G15, we find the inner radius of $R_{\rm in}\approx 10.6_{-9.3}^{+9.6}R_{\rm ISCO}$; see model 0 in Table \ref{fits}. Thus, while the model's best-fit value indicates a significantly truncated disc, it is consistent with being very close to the ISCO, as found by G15. Hereafter, the full results of the spectral fitting are given in Table \ref{fits}, while we also give some crucial values in the text.

\begin{figure}
\centering
\includegraphics[height=\columnwidth,angle=-90]{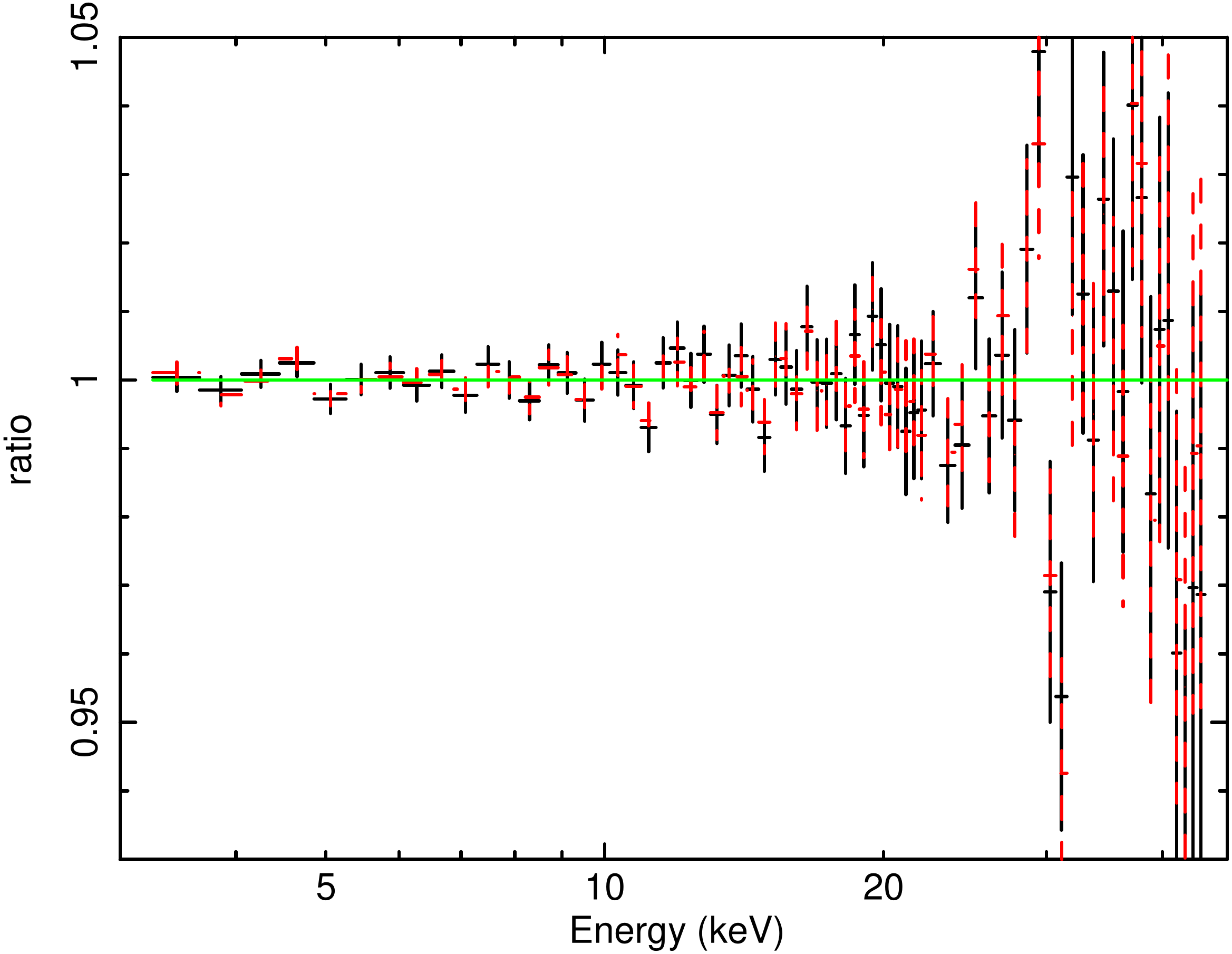}
\caption{The data-to-model ratios of the model of G15 fitted to the data sets G (black solid crosses; model 0) and A (red dashed crosses).}
\label{garcia_ratio}
\end{figure}

\begin{figure}
\centering
\includegraphics[height=\columnwidth,angle=-90]{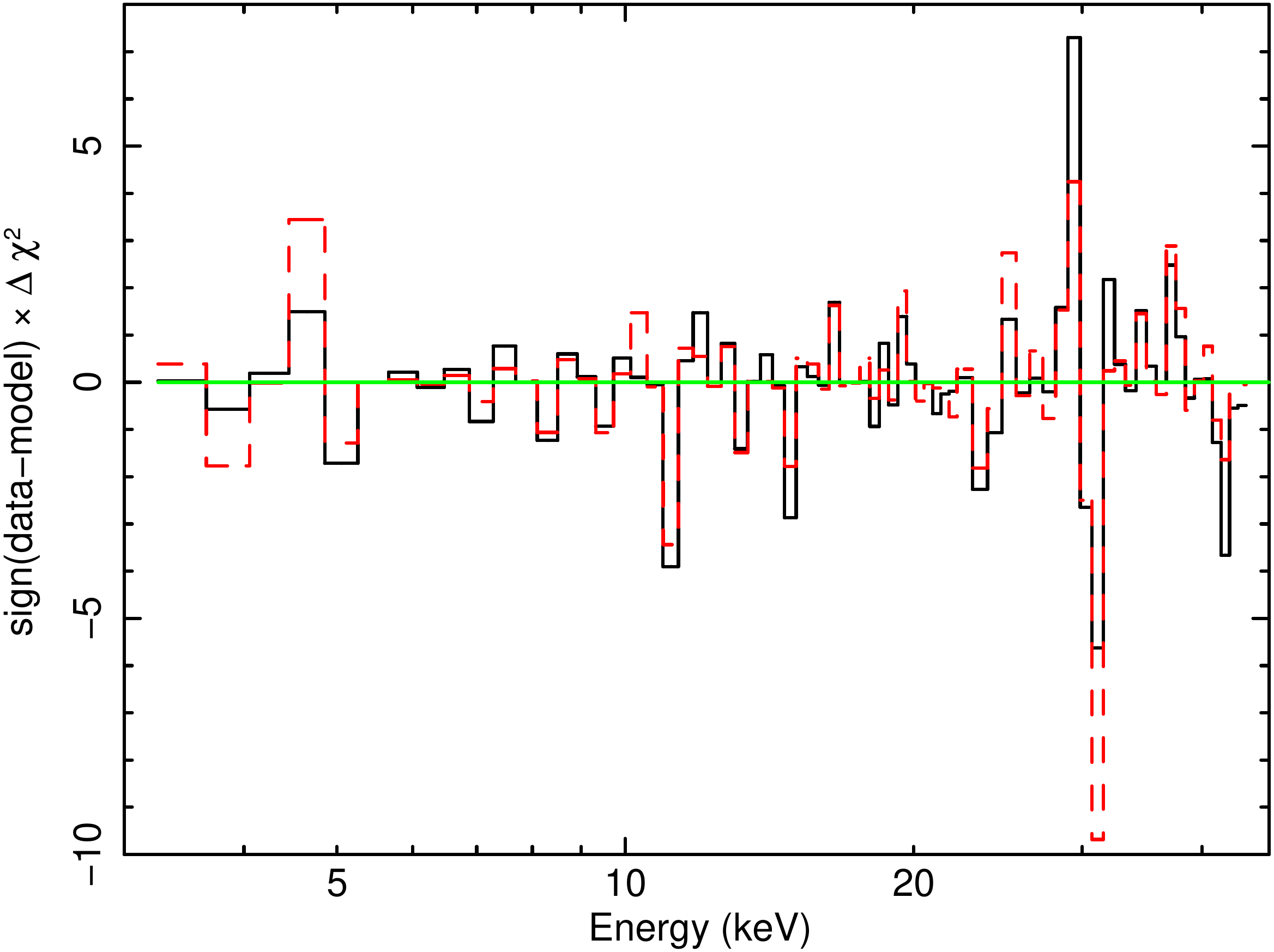}
\caption{The contributions to $\chi^2$ of the model of G15 fitted to the data set G (black solid histogram; model 0) and A (red dashed histogram).}
\label{garcia_chi}
\end{figure}

In spite of the low value of $\chi^2= 65.6$ for 61 d.o.f.\ (whose ratio we hereafter denote as $\chi_\nu^2$), we see the presence of significant residuals at energies $\ga$25 keV, as shown in Figs.\ \ref{garcia_ratio}--\ref{garcia_chi}. The origin is instrumental, due to the Xe K edge of the detector \citep{jahoda06}, but their presence does not affect the fit at lower energies, including the range of the Fe K complex. The required Fe abundance (with respect to the assumed cosmic one) is very high, $Z_{\rm Fe}=8.1^{+1.9}_{-5.5}$, where the upper limit is at the highest allowed value in the {\tt xillver} model. The reflector inclination is $i= 29_{-29}^{+31}\degr$. The contribution of the static reflection at 30 keV is about half that of the relativistically broadened reflection. The removal of the absorption line at 7.2 keV results in $\Delta\chi^2\approx +4.1$. Allowing a free line energy results in no improvement to the fit ($\Delta\chi^2\approx 0.0$).

We then fit data set A, for which the results are only slightly different. We find $R_{\rm in}\approx 14^{+2}_{-3}R_{\rm ISCO}$, $i\approx 31_{-2}^{+2}\degr$, $Z_{\rm Fe}\approx 4.1^{+5.9}_{-1.0}$, $\log_{10}\xi\approx 3.4_{-0.2}^{+0.3}$, $\chi^2_\nu\approx 64.8/61$, and other parameters similar to those found for data set G. As we can see as in Fig.\ \ref{garcia_ratio}, the model-to-data ratios for data sets G and A are very similar. The inner radius is compatible with being within $2 R_{\rm ISCO}$. For this data set, removing the absorption line results in only a slight increase in $\chi^2$, by $+1.7$. Given the similarity of the two spectra, we hereafter follow G15 and use only data set (G) obtained with their method. We note that our finding of similar-quality fits with both methods differs from that of G15, who found their fit to a summed spectrum to yield a much larger $\chi^2$ than that using their method. 

We then consider the effect of Comptonization of the reflection component. In the coronal geometry, some of the reflected emission will pass through the corona and be scattered in it. This effect has been considered by S17 using the model {\tt simplcut}\footnote{\url{http://jfsteiner.synology.me/wordpress/simplcut/}}. The main parameter of it is the scattering fraction, $f_{\rm sc}$, which is the fraction of the reflected photons that are Compton-scattered in the corona, with $1-f_{\rm sc}$ reaching the observer unmodified. Then, the scattered part of the spectrum is split between parts leaving the source and hitting the disc (where photons are removed from the model), as given by the reflection-fraction parameter of \texttt{simplcut}. In order to account for up- and down-scattering of photons out of the energy range of the PCA detector, we have extended the range of the photon energy used to calculate the models to 0.1--1000 keV with 2000 logarithmically-spaced bins. In that model, there are two options for the scattering kernel. In one, photons are scattered into an e-folded power-law distribution (with the kernel given by equation 1 of S17), the same as the incident spectrum of {\tt relxill} (see above). In the other, the Compton scattering model {\tt nthcomp} \citep*{z96} is used.

In S17, the former kernel was used, for consistency with the assumption that the incident spectrum is an e-folded power law. Those authors also replaced the incident spectrum of {\tt relxill} by {\tt simplcut(ezdiskbb)}, where {\tt ezdiskbb} is a multicolour disc blackbody model allowing for a disc truncation \citep{zimmerman05}. The resulting spectrum differs from an e-folded power law only at low energies, where the contribution of the disc blackbody is substantial. However, our data set does not show any soft excess and we have opted to keep the e-folded power law as the incident spectrum (and keep $\cal{R}$, $\Gamma$ and $E_{\rm cut}$ equal to those fitted in \texttt{relxill}). This is also consistent with the calculations of the reflection in {\tt relxill}. Thus, our model has the form {\tt tbabs[cutoffpl+simplcut(relxill)+xillver]gabs}, where the {\tt relxill} component gives now only the spectrum reflected by the e-folded power law. We have applied this model to the data set G. We have found that the best-fit scattering fraction is null, and $f_{\rm sc}=0^{+0.27}$. Thus, Comptonization of reflection does not improve the fit and does not produce significant changes to the best-fit parameters of the model of G15 as applied to the present data set and using the current version of {\tt relxill}.

\subsection{Models with relxill and a single Fe abundance}
\label{relxill}

Now, we tie the Fe abundance for the static and relativistic reflection components, and allow the former to be ionized. With this change, we find that the models no longer require the additional absorption line at 7.2 keV. Thus, our model has the form of {\tt tbabs(relxill+xillver)}. We find good fits with two kinds of models. In one, the blurred reflector is strongly ionized while the distant one is close to neutral, which is similar to the original model of G15 except for their assumptions of the separate Fe abundances and the presence of the absorption line. In the other, the blurred reflector is close to neutral while the distant one is strongly ionized. We note that both models have the same form, as given above. Thus, they actually represent two local minima of the same model. However, since their physical configurations are different while the $\chi^2$ values are similar, we consider them separately.

\begin{figure}
\centering
\includegraphics[height=\columnwidth,angle=-90]{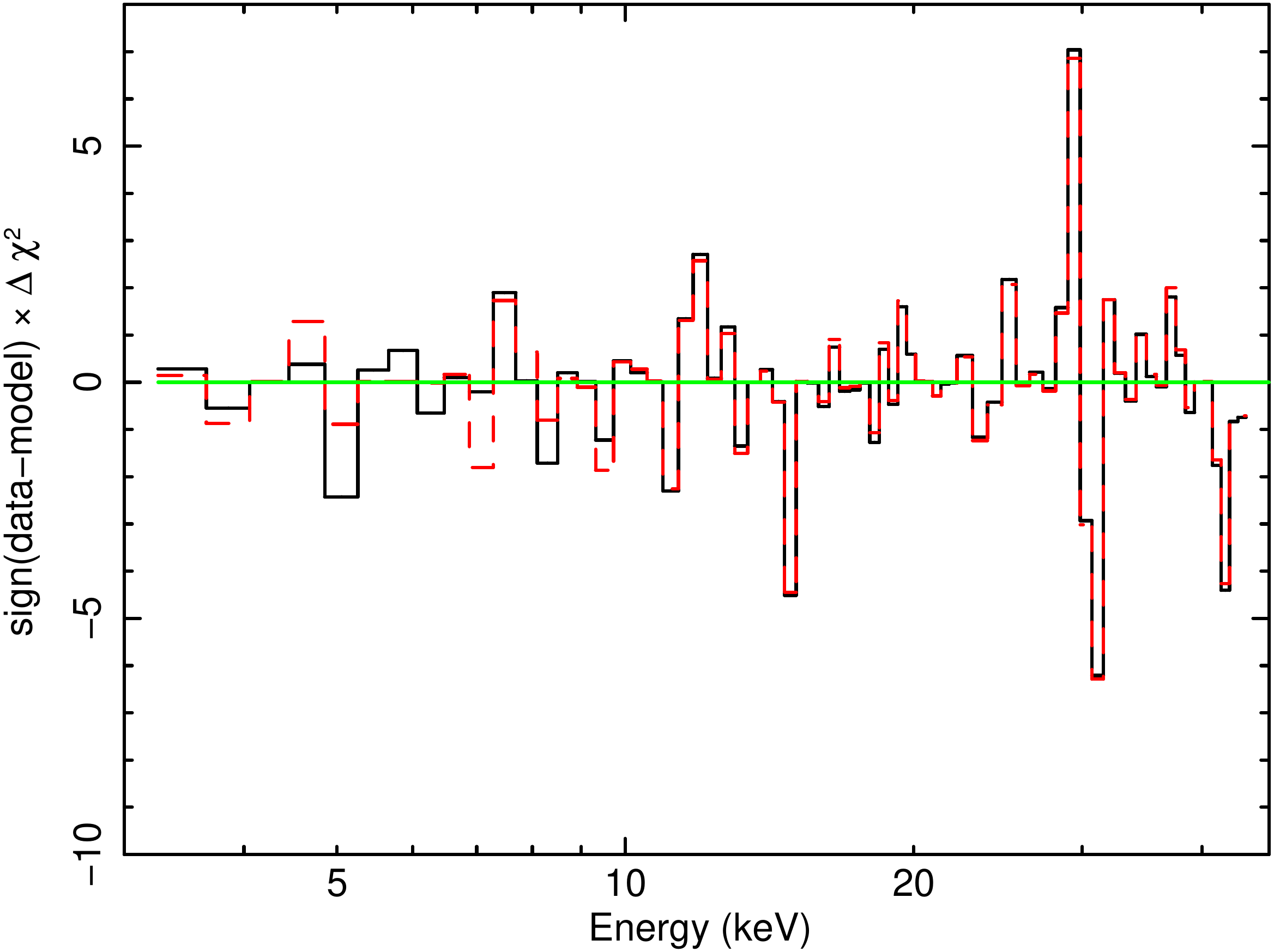}
\caption{The contributions to $\chi^2$ of the {\tt relxill}-type models with exponential cutoff. The black solid and red dashed histograms correspond to the models with the high (model 1) and low (model 2) ionization, respectively, of the close, relativistically-broadened, reflector. The vertical scales are the same as in Fig.\ \ref{garcia_chi}. }

\label{relxill_chi}
\end{figure}

In the first model, we obtain a somewhat worse fit than that in Section \ref{g15}, $\chi_\nu^2=68.7/61$, and obtain $R_{\rm in}\approx 19^{+33}_{-6}R_{\rm ISCO}$ and $Z_{\rm Fe}\approx 3.1^{+2.0}_{-0.3}$. For the second model, $\chi_\nu^2=68.3/61$, and $R_{\rm in}\approx 53^{+\infty}_{-26}R_{\rm ISCO}$, $Z_{\rm Fe}\approx 2.4^{+0.3}_{-0.2}$; see models 1 and 2, respectively, in Table \ref{fits}. Hereafter, we use the symbol $\infty$ to denote that the upper limit of $R_{\rm in}$ is approaching $R_{\rm out}$. We note that the ionization parameter of the low-ionization disc part is very weakly constrained, and allowing it to be free only marginally improves the fit. Still, here and in most of the following models, we have opted not to freeze it at $\xi=1$ in order to show the range allowed by the data. The contributions to $\chi^2$ from the two models are shown in Fig.\ \ref{relxill_chi}. We see that while the values of $\chi^2$ for these models are somewhat higher than those in Section \ref{g15}, the differences in their contributions per energy channel are very minor. We note that now the Fe abundances have lower, and much more likely, values than in the original model of G15. Also, both of our models give large truncation radii. The two models have the parameters relatively similar to each other except for the interchanged ionization parameters. This is possible because of the large fitted truncation radii, implying that the relativistic broadening is modest. There is also some difference in the strength of the reflection components. In the first model, the distant reflection component has a flux at 30 keV of about half of the flux in the relativistic component, while in the second model both reflection components' fluxes become very similar at high energies.

We then consider the effect of the electron density of the reflector. In {\tt xillver}, $n_{\rm e}=10^{15}$ cm$^{-3}$ is assumed, which is likely to be too low for accretion discs in BH binaries. The effect of the value of $n_{\rm e}$ on the reflection spectra was investigated by \citet{garcia16}. They pointed out an increase in thermal emission at low energies for a given value of $\xi$ (since then the irradiating flux is then $\propto n_{\rm e}$ and the effective temperature is $\propto n_{\rm e}^{1/4}$), which should not affect our results obtained at $\geq$3 keV. Also, the reflector temperature increases, which results in a higher ionization state. Currently, there are available models with $n_{\rm e}$ up to $10^{19}$ cm$^{-3}$, namely {\tt xillverD} and {\tt relxillD} \citep{garcia16}, while models for higher $n_{\rm e}$ are under development \citep{garcia18}. Those two models assume that the e-folding energy is fixed at 300 keV, which is within the 90 per cent confidence regime of our models 1 and 2. Thus, we fit that model to the data. We consider only the case with low ionization of the remote reflector. The fit results for this model, \#3, are given in Table~\ref{fits}. With respect to the corresponding model 1, we find a larger disc truncation radius, $R_{\rm in}\approx 55_{-34}^{+\infty} R_{\rm ISCO}$, and a higher Fe abundance, $Z_{\rm Fe}\approx 4.9_{-0.9}^{+4.1}$. The fit has a higher value of $\chi_\nu^2\approx 72.4/62$, which appears to be mostly due to the fixed value of $E_{\rm cut}$. We discuss these results in the context of other similar studies in Section~\ref{discussion}.

We have then included Compton scattering of the reflected emission in the same way as in Section \ref{g15}. We have found that in all three cases the best-fit scattering fraction was zero; see Table~\ref{fits}. 

\subsection{Models with reflkerrExp (incident e-folded power law)}
\label{reflkerr}

We now consider models of \citet{niedzwiecki18}. The differences of these models with respect to {\tt relxill} are described in detail in that paper. One difference is that the {\tt relxill} assumes the incident spectra to have the high-energy cutoff (or temperature) constant with the disc radius in the observer's frame, i.e., the incident spectra in the local frames are blueshifted with respect to that given as the model cutoff by $1+z(r)$. On the other hand, {\tt reflkerr} \citep{niedzwiecki18} assumes the incident spectrum to be constant with radius in the local frames, i.e., the observed spectrum is the sum of the local spectra redshifted by $1+z(r)$. Also, reflection in the {\tt reflkerr} model merges the detailed photoionization calculations of {\tt xillver} at low energies with the relativistically correct treatment of {\tt ireflect} \citep{mz95} at high energies.

We first consider models with incident power-law spectra with exponential cutoffs. The model is then {\tt tbabs(reflkerrExp+hreflectExp)}, where {\tt reflkerrExp} includes relativistic broadening, and {\tt hreflectExp} is the corresponding static model. Similar to our results in Section \ref{relxill}, we find two possible models, with interchanged ionization parameter. In the model with high ionization of the blurred reflector, we obtain a fit with $\chi_\nu^2\approx 69.1/61$. We find $R_{\rm in}\approx 15^{+31}_{-12}R_{\rm ISCO}$, $Z_{\rm Fe}\approx 3.9^{+0.8}_{-1.4}$; see model 4 in Table \ref{fits}. The distant reflection component contributes about half of the flux of the relativistic one at 30 keV. The bolometric flux of this model is $\approx 2.6\times 10^{-8}$ erg cm$^{-2}$ s$^{-1}$, corresponding to a luminosity of $L\approx 2.0\times 10^{38}(D/8\,{\rm kpc})^2$ erg s$^{-1}$, and $L/L_{\rm E}\approx 0.17(D/8\,{\rm kpc})^2 (M/8\,\msun)^{-1}$.

For the model with low ionization of the close reflector, we find $\chi_\nu^2\approx 69.2/61$, $R_{\rm in}\approx 58^{+\infty}_{-28}R_{\rm ISCO}$, $Z_{\rm Fe}\approx 2.6^{+0.6}_{-0.4}$, see model 5 in Table \ref{fits}. Both reflection components become almost identical at high energies. Taking into account Comptonization of the reflected radiation does not improve the fit in both cases; see the values of $f_{\rm sc}$ in Table \ref{fits}.

We can see a very good agreement between the results obtained with the current version (1.2.0) of {\tt relxill} and with {\tt reflkerrExp}. The two sets of models have almost identical parameters and the values of $\chi^2$; compare models 1 and 2 with models 4 and 5, respectively, in Table~\ref{fits}.

\begin{figure}
\centering
\includegraphics[height=\columnwidth,angle=-90]{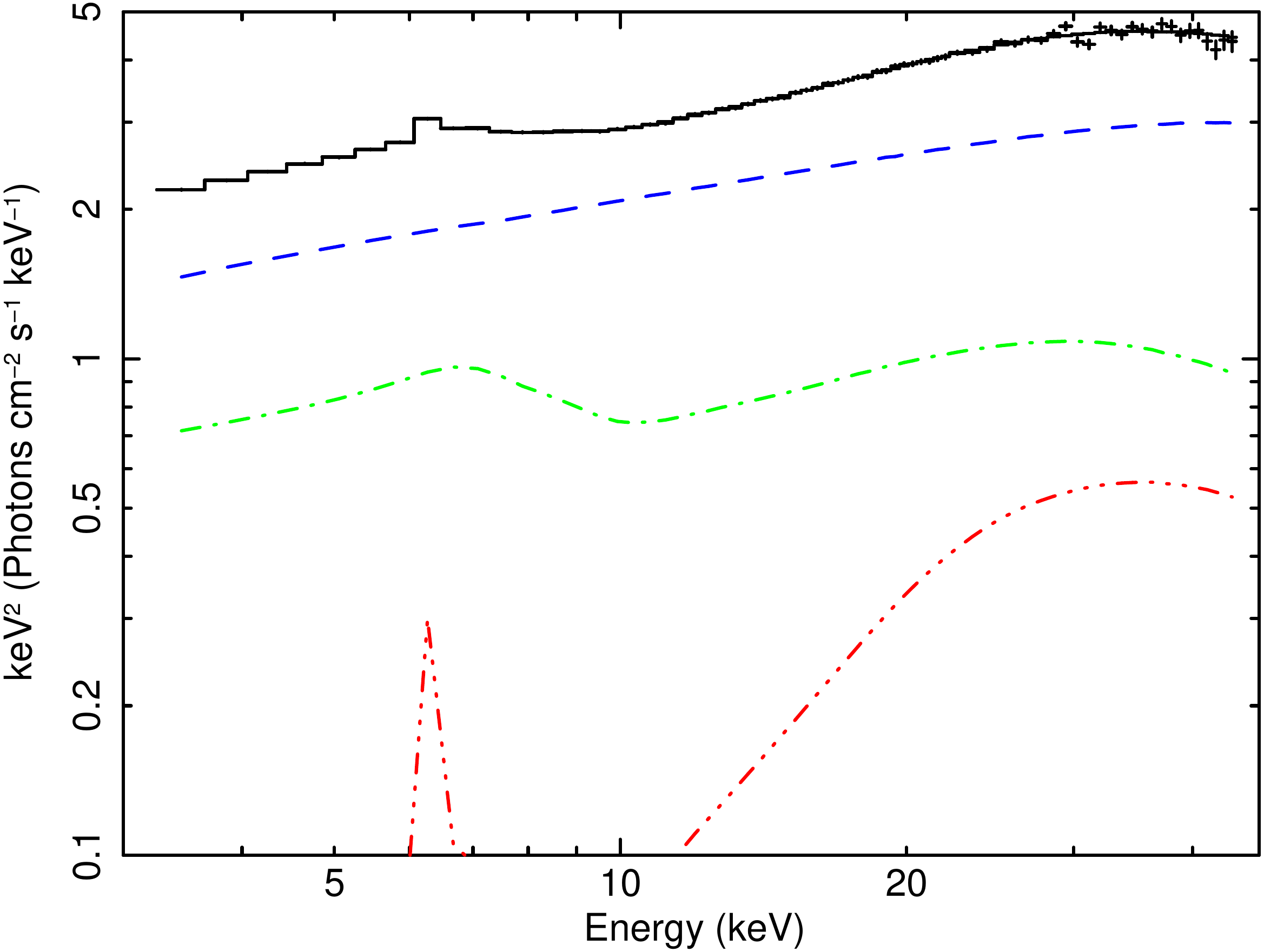}
\caption{The histogram shows the unfolded spectrum (for model 6) as fitted by thermal Comptonization (dashed blue curve) and two reflectors, an inner highly ionized one (dot-dashed green curve), and an outer weakly ionized one (triple dot-dashed red curve). The solid curve gives the total model.
}
\label{eeuf_compton}
\end{figure}

\begin{figure}
\centering
\includegraphics[height=\columnwidth,angle=-90]{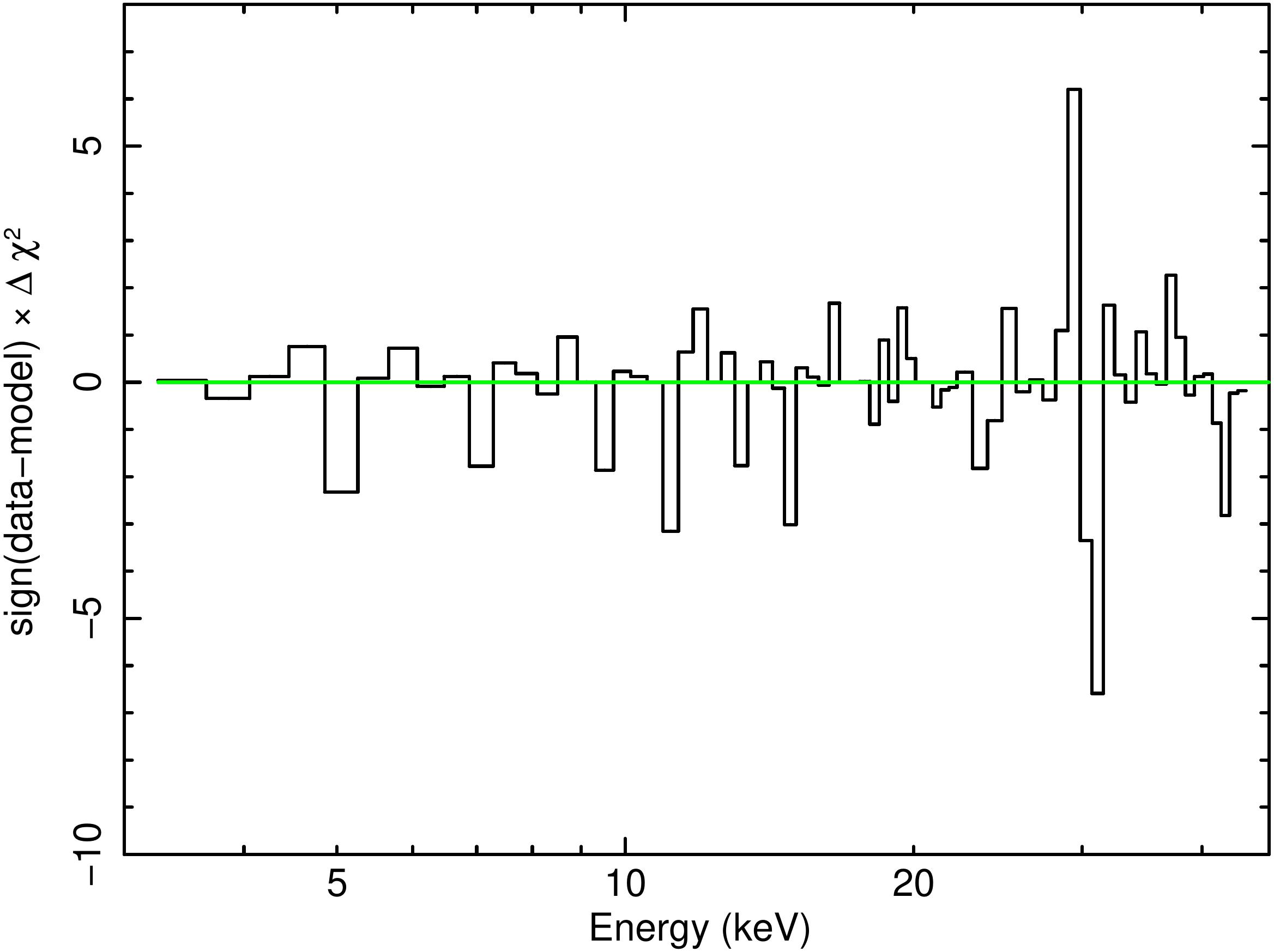}
\caption{The contributions to $\chi^2$ of the {\tt reflkerr} model (\#6) with the incident spectrum due to thermal Comptonization and high ionization of the relativistically-broadened reflector. The vertical scales are the same as in Fig.\ \ref{garcia_chi}. }
\label{reflkerr_chi}
\end{figure}

\subsection{Models with reflkerr (incident thermal Comptonization)}
\label{Compton}

\begin{figure*}
\centering
\includegraphics[height=\textwidth]{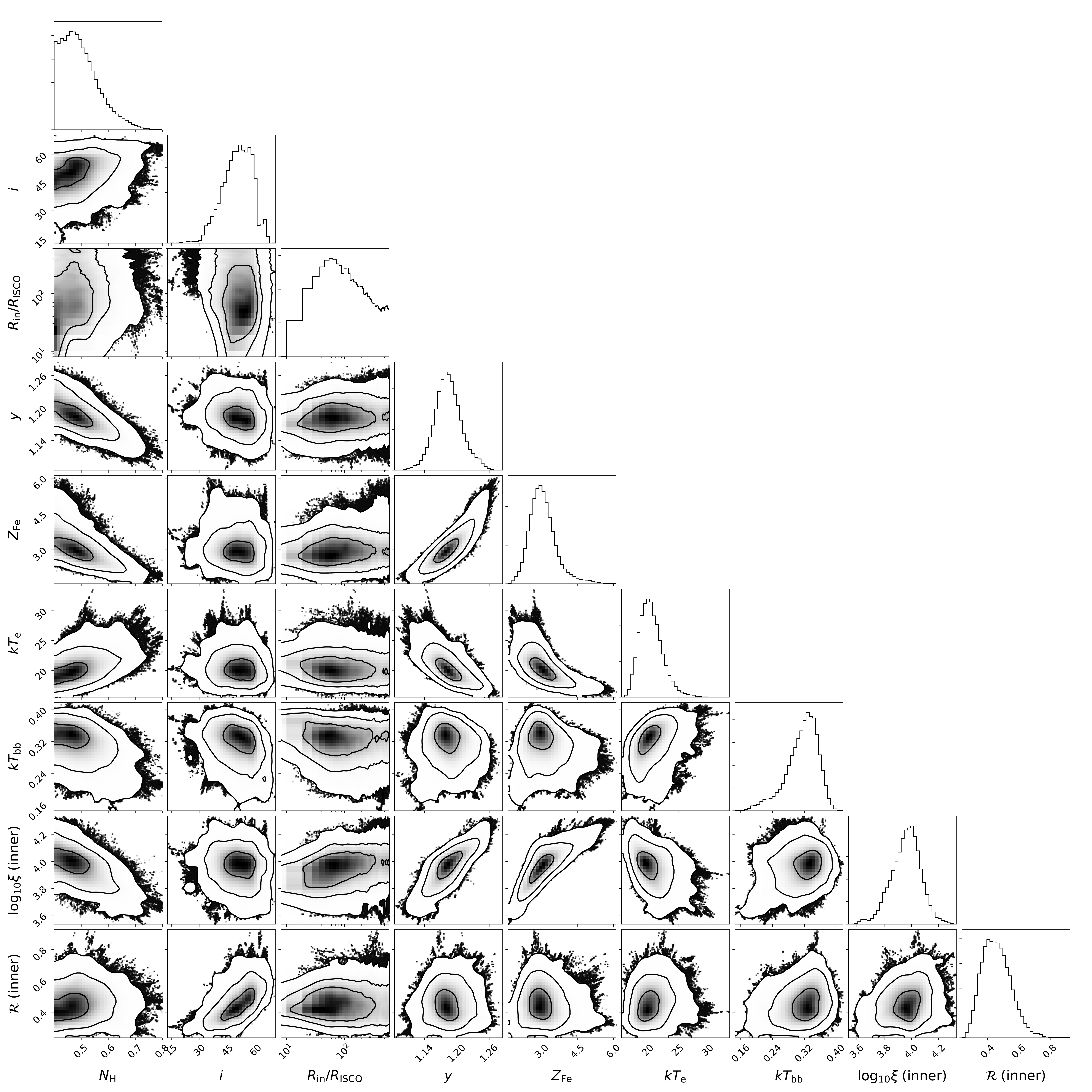}
\caption{The posterior probability distributions (proportional to the relative frequency of the fitted models) showing correlations between pairs of the parameters of the {\tt reflkerr} model (\#6), obtained using the MCMC method. The inner, middle and outer contours correspond to the 2-D significance of $\sigma=1$, 2, 3, respectively, also shown by the degree of the darkness. The rightmost panels show the probability distributions for the individual parameters (with the normalization corresponding to the unity integrated probability). See Section \ref{Compton} for discussion. }
\label{mcmc}
\end{figure*}

We then consider models of \citet{niedzwiecki18} with the incident spectrum described by the thermal Comptonization, for which they use the model {\tt compps} of \citet{poutanen96}. Our present model has the form of {\tt tbabs(reflkerr+hreflect)}. For the model with high ionization of the close reflector, we find $\chi_\nu^2\approx 62.1/61$, $R_{\rm in}\approx 47^{+\infty}_{-45}R_{\rm ISCO}$ (with the lower 90 per cent confidence limit at $R_{\rm in}\approx 1.8 R_{\rm ISCO}\approx 2.2 R_{\rm g}$) and $Z_{\rm Fe}\approx 3.3^{+1.7}_{-1.0}$; see model 6 in Table \ref{fits}. The temperature of the Comptonizing medium is $kT_{\rm e}\approx 20^{+3}_{-2}$ keV, and the temperature of blackbody seed photons is $kT_{\rm bb}\approx 0.34^{+0.04}_{-0.09}$ keV. The Compton parameter, $y\equiv 4(kT_{\rm e}/m_{\rm e}c^2)\tau_{\rm T}$, where $\tau_{\rm T}$ is the Thomson optical depth of the slab (approximating the corona), is $y\approx 1.19^{+0.05}_{-0.08}$. The distant reflection contributes about 2/3 of the flux of the relativistic one at 30 keV. In this model, in order to directly compare it to the model of G15, we have kept the ionization parameter fixed at $\xi=1$. If we allow it to be free, $\log_{10}\xi\approx 0^{+1.9}$. The unfolded spectrum and the model are shown in Fig.\ \ref{eeuf_compton} and the $\chi^2$ contributions are shown in Fig.\ \ref{reflkerr_chi}. 

The posterior probability distributions and correlations between the parameters obtained using the MCMC technique (see Section \ref{method}) are shown in Fig.\ \ref{mcmc}, obtained using the package of \citet{foreman16}. We see here rather wide probability distributions of $R_{\rm in}$, which, however, become very small for $R_{\rm in}\la 10 R_{\rm ISCO}$. The strongest correlations include the positive ones between $\xi$(inner), $Z_{\rm Fe}$ and $y$. Also, there is some positive correlation between $\cal{R}$(inner) and the inclination. Overall, we see that this model has its parameters relatively well constrained, except for the relatively wide allowed range of $R_{\rm in}$.  

Including Comptonization of reflection very slightly reduces the value of $\chi^2$, to $\chi_\nu^2\approx 62.0/60$, at $f_{\rm sc}\approx 0.30_{-0.30}^{+0.15}$, and yields very similar other parameters. The relatively low value of $f_{\rm sc}$ can be reconciled with the relatively large $\tau_{\rm T}$ of the model in the geometry of a central hot flow surrounded by a truncated disc, in which relatively few reflected photons return into the hot flow, see, e.g., \citet*{z99}, \citet*{poutanen18}.

S17 also considered models without the presence of a distant reflector. Here, we confirm their conclusion that such models give a much worse description of the data using our model 6. If we do not include the static reflection component, we obtain $\Delta\chi^2\simeq +22$ for one less d.o.f., which corresponds to the probability of the fit improvement by adding that component being by chance of $2\times 10^{-5}$ (using the F-test).

In the case with low ionization of the close reflector, we also find a very good model, for which $\chi_\nu^2\approx 63.8/60$, $R_{\rm in}\approx 100^{+\infty}_{-62}R_{\rm ISCO}$, $Z_{\rm Fe}\approx 2.9^{+0.5}_{-0.3}$ and $i\approx 18^{+57}_{-6}\degr$. The high-ionization, distant reflection actually dominates in this model, with its flux at 30 keV being higher by $\approx 1.7$ than that of the close reflection. Including Comptonization of reflection (using {\tt simplcut} with the {\tt nthcomp} scattering kernel) improves the fit only marginally. For the sake of the simplicity of the presentation, this model is not included in Table~\ref{fits}.

\section{Discussion and conclusions}
\label{discussion}

Our main finding is that X-ray spectra of the hard state of \source from the PCA can be fitted with similar statistical quality with models allowing significantly different disc truncation radii, namely with $R_{\rm in}$ either close to or much larger than $R_{\rm ISCO}$. Still, all of the fitted models prefer the latter at their best-fit values. This is the case ($R_{\rm in}\approx 11 R_{\rm ISCO}$) even for the original model (\#0 in Table \ref{fits}) of G15 fitted to our average PCA spectrum with the current version of the {\tt relxill} software. That model, however, requires the presence of a 7.2 keV absorption line, separate Fe abundances for the two reflectors, and a high value of the Fe abundance for the relativistic reflector.

While the presence of an absorption line at 7.2 keV is in principle possible, it has not been found in other observations of \source. In systematic studies of BH X-ray binaries, \source shows the same properties as systems without disc-wind absorption lines. In particular, the shape of the track on the hardness-count rate diagram of \source favours a relatively low inclination \citep{munoz13}. Also, high-ionization Fe K absorption lines in the soft state, which trace disc winds, have not been detected in \source \citep{ponti12}. Then, while the assumption of strongly different Fe abundances in the two reflectors can be motivated by our lack of knowledge of the true model of the accretion flow, it does not have a direct physical interpretation. 

On the other hand, we have found alternative models (\#1--6 in Table \ref{fits}), which rule out $R_{\rm in}\approx R_{\rm ISCO}$ at 90 per cent confidence. In particular, if we impose a common Fe abundance in both reflection components and keep the assumption of the exponential cutoff of the incident spectrum (models 1--5), we find $R_{\rm in}\ga 6 R_{\rm ISCO}$, using either {\tt relxill} or {\tt reflkerrExp}. The models no longer require the 7.2 keV absorption line and their fitted Fe abundances are relatively moderate, $Z_{\rm Fe}\ga 2$. Still, they have somewhat higher value of $\chi^2$ than the model following the assumptions of G15, with $\Delta \chi^2\approx 2$--3. Also, those models in the variant with high ionization of the close reflector (\#1, 3, 4, which appear more likely; see below) yield the disc inclinations close to face-on, which are only marginally consistent with the constraint on the binary inclination of $37\degr\leq i_{\rm b}\leq 78\degr$ \citep{heida17}. Since those models imply the presence of a truncated disc with $R_{\rm in}\gg R_{\rm ISCO}$, the disc axis is likely to be aligned with the binary axis rather than the BH spin axis. We note that S17 argued that disc truncation in \source requires a highly super-Eddington accretion rate. However, that argument is incorrect since it considers only the viscously generated soft seed photons but neglects the abundant soft photons from reprocessing of the hot-flow emission in the irradiated disc, as discussed in detail in \citet{poutanen18}. Both the viscously generated photons and those due to re-emission of the irradiating flux absorbed in the disc form then a soft quasi-blackbody spectral component around the disc inner radius. This component corresponds to the blackbody seed photons in model 6. 

Our overall best-fit model (\#6 in Table \ref{fits}), with the lowest value of $\chi_\nu^2\approx 62/61$ and the by far highest Akaike likelihood, has a physical thermal-Comptonization primary continuum rather than a phenomenological e-folded power law. The best fit value of the truncation radius is several tens of $R_{\rm ISCO}$, and $R_{\rm in}\ga 2 R_{\rm ISCO}$ within the 90 per cent confidence limits. The Fe abundance is the same for both reflectors, and moderate, $Z_{\rm Fe}\approx 3$, and there is no absorption line at 7.2 keV required. This model also has the best-fit inclination, $i\approx 49^{+34}_{-26}\degr$, in good agreement with the constraint of \citet{heida17}. We stress this is the only model among those considered that has a moderate Fe abundance, an inclination in agreement with that measured for the binary and the ionization of the distant reflection lower than that of the inner disc (which appears more likely; see below). On the other hand, only some, but not all, of those desired features appear in other models considered.

The primary continuum in this thermal-Comptonization model differs quite significantly from an e-folded power law at high energies. The difference between the two types of spectra is illustrated, e.g., in fig.\ 5b of \citet{z03}. The Comptonization spectrum in that case has $kT_{\rm e}=25$ keV, close to our best fit value of $20^{+3}_{-2}$ keV. We see in that figure that the Comptonization spectrum has an approximately single-power law shape up to $\approx$50 keV, and then it shows a sharp cutoff. On the other hand, an e-folded power law features a gradual attenuation, visible already at $E\ll E_{\rm cut}$. In the example shown in \citet{z03}, an e-folded power law with $E_{\rm cut}=150$ keV lies below the Comptonization spectrum already by $E\ga 10$ keV. This effect explains the mismatch between the values of $kT_{\rm e}$ and $E_{\rm cut}$ that we found, with $E_{\rm cut}\gg kT_{\rm e}$ fitted here to the same spectrum at $E\leq 45$ keV. 

For all of our models, we have also taken into account Comptonization of the reflected component. However, we have found this effect to be minor. In model 0 (which follows the assumptions of G15), the reflection fraction is ${\cal R}\approx 0.06$ and the fraction of the reflection photons scattered in the corona is $\la 0.3$; such values rule out a corona above a disc extending to the ISCO. On the other hand, the relatively low values of $\cal R$ in our other models are consistent with truncation and the primary continuum originating partly in a corona and partly in a hot flow at $R<R_{\rm in}$ \citep{z99}. Our physical (and statistically best) thermal Comptonization model is also compatible with the energy balance constraint; see, e.g., the recent study of \citet{poutanen18}. 

We have also investigated the effect of increasing the reflector density. Using the model of \citet{garcia16} with $n_{\rm e}=10^{19}$~cm$^{-3}$ (\#3 in Table \ref{fits}), we have found an increase of both the truncation radius and the Fe abundance with respect to the corresponding model with an exponential cutoff and $n_{\rm e}=10^{15}$ cm$^{-3}$. In particular, $Z_{\rm Fe}$ increased from $\approx$3 to $\approx$5 at the best fit.

The increase in $Z_{\rm Fe}$ we found is surprising, given the results of \citet{tomsick18}. They have studied the effect of changing the reflector density for the case of a hard/intermediate spectrum of Cyg X-1 from \suzaku and \nustar. In the case of free Fe abundance, coronal geometry, and $n_{\rm e}= 10^{15}$ cm$^{-3}$, they obtained, using {\tt relxill}+{\tt xillver}, an extreme abundance of $Z_{\rm Fe}\sim$10 and a very low disc inner radius of $R_{\rm in}=1.33_{-0.06}^{+0.03} R_{\rm ISCO}$. Then, they used the reflection model {\tt reflionx\_hd} of \citet{ross07}, which assumes $Z_{\rm Fe}=1$ but allows for a free reflector density. The best fit in that case was obtained for $n_{\rm e}\approx 4\times 10^{20}$ cm$^{-3}$, a much larger truncation radius, $R_{\rm in}\approx 7.3_{-1.9}^{+4.6} R_{\rm ISCO}$, and at a much lower value of $\chi^2$ than in the previous case. Thus, the effects of allowing a high density in that case are an increase of the truncation radius and obtaining a good fit at the solar Fe abundance. While the former is in agreement with our finding, the latter effect is opposite. The reason for this disagreement is unclear; it may be related to a difference in the treatment of atomic processes in the reflecting/reprocessing medium between the {\tt reflionx\_hd} and {\tt relxill} models.

We note that the actual characteristic value of $n_{\rm e}$ of the reflecting medium is uncertain. \citet{garcia16} estimated it using the radiation-pressure dominated disc solution including the effect of coronal dissipation of \citet{sz94}. However, this density is averaged over the disc height, and it is close to that in the disc midplane. The corresponding Thomson optical depth of that solution is $\gg 10$, and the reflecting surface layer has an optical depth of several and a much lower average density, given by the vertical hydrostatic equilibrium. Furthermore, the actual density is a function of both the radius and the depth within the disc.

A possible check on the self-consistency of an approximate solution is using the definition of $\xi$, equation (\ref{xi}). The irradiating flux in a coronal geometry can be expressed as the ratio of the luminosity to the emitting area, where the latter can be estimated as $x R_{\rm in}^2$, where the constant $x\sim 1$ expresses our (large) uncertainty about the area. In the case of Cyg X-1, $L\approx 1.8 \times 10^{37}$ erg s$^{-1}$, $R_{\rm in}\sim 10 R_{\rm g}$ \citep{tomsick18} and $M\approx 15\msun$ \citep{orosz11}. The fitted ionization parameter was $\xi\approx 2000$ erg cm$^{-2}$ s$^{-1}$, which implies $n_{\rm e}\approx (2/x)\times 10^{20}$ cm$^{-3}$, rather close to the density fitted in that work. In the case of our observation of \source and model 3, $L\approx 2 \times 10^{38}$ erg s$^{-1}$, $\xi\approx 5000$ erg cm$^{-2}$ s$^{-1}$, $R_{\rm in}\sim 50 R_{\rm ISCO}$, which yields $n_{\rm e}\approx (2/x)\times 10^{19}$ cm$^{-3}$. This implies that only model 3 can be considered as approximately self-consistent. However, given that the current version of the {\tt xillverD} model allows for only for one option of the high-energy cutoff, we could not consider other options with it, in particular that with a thermal Comptonization primary spectrum.

In our analysis, we found that the data allow the ionization of the outer reflector to be higher than the inner one. This effect is due to the relatively modest relativistic effects in the inner reflector, which then allows for its interchange with the outer, static, one. Since $\xi\propto F_{\rm irr}/n_{\rm e}$, it is not a priori obvious that the surface layers of the outer disc are weakly ionized, given that they are irradiated by the strong X-ray emission from the central source, with the disc likely to be flared. Still, those solutions (\#2, 5 in Table \ref{fits}) appear less likely. We give them for the sake of the completeness of the presentation of our analysis results.

We stress that our most physically-motivated case uses a one-zone thermal Comptonization model. In the case of the hard state of the BH binary Cyg X-1, \citet{axelsson18} found that the X-ray variability properties require the presence of three separate Comptonization components with different spectral slopes and temperatures. \citet{yamada13} obtained a similar conclusion from analysing the broad-band variability in Cyg X-1. \citet{mahmoud18a,mahmoud18b} modelled the spectral and timing properties of Cyg X-1, finding that rather complex models are required. In \source, observational evidence for two Comptonization zones has been recently reported from the full spectral-timing modelling of one \xmm hard-state observation of the source \citep{mahmoud19}. Given that result, and the overall similarity of the spectral and timing properties of Cyg X-1 and \source, our one-zone Comptonization modelling appears to be too simple to describe the actual accretion flow. However, since the \xte\/ data studied here cover the range of $\ga$3 keV only, the second, soft, Comptonization component in the model of \citet{mahmoud19} (see their fig.\ 4) contributes negligibly to the range fitted by us, and it does not affect the validity of our results.

We can compare our models to those that \citet{basak16} fitted to their EPIC-pn spectrum 7, in particular for their model 2(ii), which includes a static reflection component. That fit yields $R_{\rm in}=19.5^{+15.0}_{-8.0}R_{\rm g}$ and $\Gamma\approx 1.67^{+0.02}_{-0.02}$, which are compatible with our results. The main differences are the high ionization of the outer static reflector, about the same as that of the inner one, and the Fe abundance, which they find to be about solar, with $Z_{\rm Fe}=0.95^{+0.07}_{-0.06}$. \citet{basak16} also fitted a simultaneous PCA spectrum, and noticed a mismatch between the EPIC-pn and PCA calibration in the Fe K region; see their fig.\ 8.

We note that a PCA spectrum of \source in the hard spectral state has been used to  test an alternative GR theory as well as to measure the BH spin \citep{wang_ji18a}, assuming that the disc extends to the ISCO (see their table II). In light of our results, that test appears to be highly uncertain.

We stress that we have found a very good agreement between the models using the present version of {\tt relxill} and {\tt reflkerrExp}. They yield almost identical values of $R_{\rm in}$ and other parameters at very similar $\chi^2$; see Table \ref{fits}. 

Concluding, our spectral fitting results support the truncated disc paradigm for the hard state \citep{done07}, but allow the reflecting disc to extend to within about two ISCO radii within the 90 per cent confidence limit. Still, our results are based on a single spectrum (though with a very large count number of $10^7$). Stronger constraints can be obtained while simultaneously fitting several data sets (as G15 did for their particular model).

\section*{Acknowledgments}
We thank the referee for valuable comments, Javier Garc{\'{\i}}a for discussions and explanations regarding the spectral analysis performed in G15, Jorge Casares and Jean-Pierre Lasota for discussions about disc inner radii in the quiescent state, Piotr Lubi{\'n}ski for discussion about the Akaike formalism, and Michael Parker for advice on the MCMC plotting. This research has been supported in part by Polish National Science Centre grants 2013/10/M/ST9/00729, 2015/18/A/ST9/00746, 2016/21/B/ST9/02388 and 2016/21/P/ST9/04025.

\label{lastpage}

\end{document}